\documentclass[english,aps,prl,superscriptaddress,twocolumn]{revtex4-2}
\usepackage[T1]{fontenc}
\usepackage[latin9]{inputenc}
\usepackage{amsmath}
\usepackage{graphicx}
\usepackage{babel}
\begin{document}
\title{Nonequilibrium entropy from density estimation}
\author{Samuel D. Gelman}
\affiliation{School of Chemistry, Tel Aviv University, Tel Aviv 6997801, Israel}
\author{Guy Cohen}
\email{gcohen@tau.ac.il}

\affiliation{School of Chemistry, Tel Aviv University, Tel Aviv 6997801, Israel}
\affiliation{The Raymond and Beverley Sackler Center for Computational Molecular
and Materials Science, Tel Aviv University, Tel Aviv 6997801, Israel}
\date{\today}
\begin{abstract}
Entropy is a central concept in physics, but can be challenging to
calculate even for systems that are easily simulated. This is exacerbated
out of equilibrium, where generally little is known about the distribution
characterizing simulated configurations. However, modern machine learning
algorithms can estimate the probability density characterizing an
ensemble of images, given nothing more than sample images assumed
to be drawn from this distribution. We show that by mapping system
configurations to images, such approaches can be adapted to the efficient
estimation of the density, and therefore the entropy, from simulated
or experimental data. We then use this idea to obtain entropic limit
cycles in a kinetic Ising model driven by an oscillating magnetic
field. Despite being a global probe, we demonstrate that this allows
us to identify and characterize stochastic dynamics at parameters
near the dynamical phase transition.
\end{abstract}
\maketitle

\paragraph{}

Evaluating entropy of statistical mechanical systems from simulations
or experimental snapshots is a challenge that has drawn attention
repeatedly for more than half a century. In equilibrium, this issue
was largely resolved by the introduction of celebrated and widely
used concepts like thermodynamic integration \citep{hansen_phase_1969}
and later the Wang--Landau method \citep{wang_efficient_2001}. On
the other hand, entropy is of particular interest in nonequilibrium
situations, where most state variables are ill-defined \citep{cross_pattern_1993,martiniani_quantifying_2019}.
Here, despite decades of study, the question of how entropy can be
generically and efficiently evaluated has never been fully answered
\citep{binder_monte_1985,beirlant_nonparametric_1997,zu_information_2020,sorkin_universal_2023,fraenkel_information_2024}.

The problem is succinctly stated as follows. Suppose we can simulate
the statistics of a system. In all that follows, simulating a system
may be considered equivalent to experimentally realizing and fully
characterizing it. The ability to simulate implies that we can access
a set of $N$ sample configurations $\left\{ \mathbf{x}_{i}\right\} $,
with index $i\in\left\{ 1,\ldots,N\right\} $, drawn from some (unknown)
probability distribution $P\left(\mathbf{x}\right)$ describing the
system's behavior at some moment in time. Given these samples, we
can estimate expectation values of various properties of the system.
For example, given the energy of a configuration $E\left(\mathbf{x}\right)$,
the mean energy is
\begin{equation}
\left\langle E\right\rangle =\int\mathrm{d}\mathbf{x}P\left(\mathbf{x}\right)E\left(\mathbf{x}\right)=\sum_{i=1}^{N}\frac{E\left(\mathbf{x}_{i}\right)}{N}+O\left(\frac{1}{\sqrt{N}}\right).\label{eq:energy_from_samples}
\end{equation}
Importantly, eq.~\eqref{eq:energy_from_samples} can be evaluated
without knowing $P\left(\mathbf{x}\right)$. On the other hand, analogous
considerations for the Von Neumann entropy (or similarly the free
energy) give
\begin{equation}
\begin{aligned}\left\langle S\right\rangle  & =\frac{1}{N}\sum_{i=1}^{N}\ln\left[P\left(\mathbf{x}_{i}\right)\right]+O\left(\frac{1}{\sqrt{N}}\right),\end{aligned}
\label{eq:entropy_from_von_neumann}
\end{equation}
and knowledge of $P\left(\mathbf{x}\right)$ is expressly required.
In equilibrium, some progress can be made from knowing that $P\left(\mathbf{x}\right)\sim e^{-\frac{E\left(\mathbf{x}\right)}{k_{B}T}}$;
but for generic nonequilibrium situations we have no a priori knowledge
of the distribution.

\paragraph{}

Numerous routes around this have been proposed in the literature,
from the beautiful early work of Ma on coincidence counting in 1981
\citep{ma_calculation_1981} to increasingly sophisticated recent
approaches based on compression algorithms \citep{avinery_universal_2019,martiniani_quantifying_2019}
and machine learning models for mutual information between subsystems
\citep{nir_machine-learning_2020}. Other approaches are approximate,
for example relying on structure factors \citep{ariel_inferring_2020,sorkin_resolving_2023}
or making other assumptions \citep{panda_non-parametric_2023}. Rapidly
increasing interest in applying machine learning techniques to a variety
of physical problems \citep{carleo_machine_2019,atanasova_stochastic_2023,bernheimer_determinant_2024}
makes many of these directions appealing. Despite these continuing
advances, existing methodologies are either slow to converge in the
number of samples; intrinsically biased; or too computationally expensive
to be applied to large systems without additional assumptions.

\paragraph{Entropy from density estimation.}

Here, we propose a surprisingly straightforward and efficient solution
based on first solving a seemingly harder problem: density estimation.
In machine learning terms, density estimation is the task of parametrizing
an optimal probability distribution $P_{\boldsymbol{\vartheta}}\left(\mathbf{x}\right)$
from a set of samples $\mathbf{x}_{i}$ taken from an unknown distribution
$P\left(\mathbf{x}\right)$; here $\boldsymbol{\vartheta}$ is a set
of parameters. There is no unique criterion for what should be considered
optimal, but a common approach is to select $\boldsymbol{\vartheta}$
so as to minimize the Kullback--Leibler divergence (KLD) between
$P_{\boldsymbol{\vartheta}}\left(\mathbf{x}\right)$ and the prior
distribution $P_{\mathrm{Prior}}\left(\mathbf{x}\right)\equiv\frac{1}{N}\sum_{i=1}^{N}\delta\left(\mathbf{x}-\mathbf{x}_{i}\right)$.
Given a model distribution $P_{\boldsymbol{\vartheta}}\left(\mathbf{x}\right)$
and samples $\mathbf{x}_{i}$, eq.~\eqref{eq:entropy_from_von_neumann}
can then be approximated by
\begin{equation}
\left\langle S\right\rangle \approx\frac{1}{N}\sum_{i=1}^{N}\ln\left(P_{\boldsymbol{\vartheta}}\left(\mathbf{x}_{i}\right)\right).\label{eq:entropy_from_model_distribution}
\end{equation}

Perhaps the simplest density estimation algorithm is the division
of space into bins and the construction of a normalized histogram.
Another simple approach, known as kernel density estimation, involves
convolving $P_{\mathrm{Prior}}\left(\mathbf{x}\right)$ with, e.g.,
a Gaussian function $P_{\boldsymbol{\vartheta}}\left(\mathbf{x}\right)={\displaystyle \frac{1}{\sqrt{2\pi\sigma^{2}N^{2}}}\sum_{i=1}^{N}e^{-\left(\frac{\mathbf{x}-\mathbf{x}_{i}}{4\sigma}\right)^{2}}}$.
The lone parameter $\boldsymbol{\vartheta}=\sigma$ then expresses
a trade-off between the smoothness and resolution of the resulting
distribution. However, histograms and kernel methods do not scale
well to high dimension. Recent interest in applying density estimation
to extremely high-dimensional problems like image processing has resulted
in a variety of novel and highly accurate generative models based
on artificial neural networks \citep{rezende_variational_2015,balle_density_2016,van_den_oord_conditional_2016,kingma_improved_2016,salimans_pixelcnn_2017,papamakarios_masked_2017,liu_density_2021,papamakarios_neural_2019,bond-taylor_deep_2021,kingma_variational_2021}.

Generative models earn their name from their ability to generate samples
drawn from the distribution they describe. In the context of statistical
mechanics, this has been taken advantage of to not only characterize---but
rather to effectively simulate---equilibrium models, by minimizing
the variational free energy of generated samples \citep{wu_solving_2019,nicoli_asymptotically_2020,singh_conditional_2021,damewood_sampling_2022,ma_message_2024}.
This approach requires less ad hoc simulation and analysis than traditional
approaches like Wang--Landau, but is much more expensive, and in
practice it has only been used for small systems. In particular, ref.~\citep{nicoli_asymptotically_2020}
solved a $24\times24$ square Ising model, whereas ref.~\citep{nir_machine-learning_2020}
reached $64\times64$; for comparison, Wang and Landau demonstrated
their approach on the $256\times256$ case two decades earlier \citep{wang_efficient_2001}.

\paragraph{}

In what follows, we first demonstrate that by leveraging standard
simulations to generate samples and using generative density estimation
to learn the distribution, system sizes comparable with those treatable
by the Wang--Landau algorithm become easily accessible. We then show
that nonequilibrium systems become just as tractable, and use our
approach to investigate the dynamics of entropy in a nonequilibrium
Ising model driven by an oscillating magnetic field. There, we study
the entropic limit cycle at the boundary between the ferromagnetic
and paramagnetic phases, showing that it reveals a nonequilibrium
critical regime with long-lived hysteretic behavior.

\paragraph{Autoregressive ansatz.}

While the evaluation of entropy from density estimation is a completely
general idea, it is useful to limit our attention to physical models
for which configurations may be expressed as two dimensional images.
To be concrete, we assume $\left(\mathbf{x}_{i}\right)^{nm}=\sigma_{i}^{nm}$,
where $n\in\left\{ 1,2,\ldots,L_{1}\right\} $ and $m\in\left\{ 1,2,\ldots,L_{2}\right\} $
are spatial indices; $\sigma_{i}^{nm}$ can take on any of some discrete
set of values; and $L_{1}$ and $L_{2}$ determine the system's size.
For example, in the Ising model on an $L\times L$ square lattice,
$\sigma\in\left\{ -1,1\right\} $ and $L_{1}=L_{2}=L$.

Given the assumptions above, we employ a highly successful approach
to density estimation known as PixelCNN++ \citep{oord_pixel_2016,van_den_oord_conditional_2016,salimans_pixelcnn_2017}.
This is an autoregressive model, meaning that to make the joint distribution
$P\left(\mathbf{x}\right)$ tractable, it is decomposed using the
chain rule into a product of conditionals:
\begin{equation}
P_{\boldsymbol{\vartheta}_{1},\ldots,\boldsymbol{\vartheta}_{N}}\left(\mathbf{x}\right)=\prod_{\ell=1}^{N}P_{\boldsymbol{\vartheta}_{\ell}}\left(\left.x^{n_{\ell}m_{\ell}}\right|\left\{ x^{n_{\ell^{\prime}}m_{\ell^{\prime}}},\ell^{\prime}<\ell\right\} \right).
\end{equation}
Here, the indices $\ell$ and $\ell^{\prime}$ define a 1D ordering
over the spatial indices $n$ and $m$ (i.e. over spins in an Ising
model) and $N=L_{1}\cdot L_{2}$. The conditionals $P_{\boldsymbol{\vartheta}_{\ell}}$
themselves are expressed in terms of an efficient neural network ansatz
that allows for their rapid training in parallel, and the set of spins
on which each spin's conditional depends is effectively limited to
simplify training \citep{salimans_pixelcnn_2017}. Despite its intrinsically
1D nature, it turns out that a sequential scan over lines in 2D images
often works very well in describing the distributions of complicated
ensembles \citep{bond-taylor_deep_2021}.

\paragraph{Equilibrium Ising model.}

\begin{figure}
\includegraphics{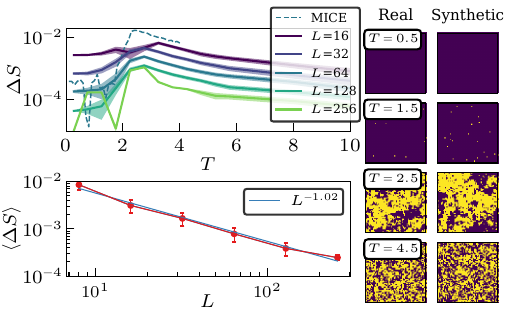}

\caption{(upper left panel) Difference in entropy from exact result at different
system sizes $L$, as a function of temperature $T$. For comparison,
the dashed line is the $L=64$ MICE data from ref.~\citep{nir_machine-learning_2020}.
(lower left panel) Temperature-averaged entropy difference, with standard
deviation as error bars. (right panels) Monte Carlo samples at different
temperature (left column), next to samples drawn from the generative
ansatz at the same temperatures (right column).\label{fig:equilibrium_entropy}}
\end{figure}

We begin by benchmarking our approach against known results for the
ferromagnetic Ising model on a square lattice with toroidal boundary
conditions, given by
\begin{equation}
E\left(\mathbf{x}\right)=E\left(\boldsymbol{\sigma}\right)=-J\sum_{\left\langle nm,n^{\prime}m^{\prime}\right\rangle }\sigma^{nm}\sigma^{n^{\prime}m^{\prime}}+\underset{nm}{\sum}h\sigma^{nm},
\end{equation}
where the sum in the first term is taken over nearest neighbors. Here,
we set $J=1$ and employ it as our unit of energy. The system is simulated
using a standard Monte Carlo procedure, with the Metropolis--Hastings
algorithm\citep{metropolis_equation_1953} used at higher temperatures
and the Wolff algorithm \citep{wolff_collective_1989} used at critical
and low temperatures. We generate 10,000 independent training samples
and attempt to minimize a loss function comprising the KLD of $P_{\boldsymbol{\vartheta}_{1},\ldots,\boldsymbol{\vartheta}_{N}}\left(\mathbf{x}\right)$
with the prior. Training continues until the loss associated with
1,000 additional test samples reaches a plateau. A final set of samples
is then used to estimate the entropy using Eq.~\eqref{eq:entropy_from_model_distribution},
and a standard error is obtained from five independent calculations
of this type.

The top left panel of Fig.~\ref{fig:equilibrium_entropy} shows the
error in our estimate for the entropy as a function of temperature,
at several system sizes ranging from $L=16$ to $L=256$. Note that
this is obtained from comparison with the exact result at finite system
size \citep{ferdinand_bounded_1969}, not with the thermodynamic limit
\citep{onsager_crystal_1944}; other than statistical uncertainty,
which is easily accounted for, this error is therefore entirely due
to the bias introduced by the machine learning ansatz and its optimization,
and is always positive. For comparison, we also reproduce the MICE
result from Ref.~\citep{nir_machine-learning_2020} for $L=64$ (dashed
line). The latter was obtained with 5,000 samples, but used data augmentation
techniques to increase the effective number available.

We use the same number of samples for all system sizes, but each sample
contains $L^{2}$ bits of information; therefore, more data was available
at large system sizes, and as a result accuracy (generally) improves.
On the other hand, one might expect larger systems to be harder to
characterize---this, especially in the critical regime, where large
length scales appear and errors are maximal. To gain some understanding
of the method's behavior, we plot the temperature-averaged entropy
as a function of system size in the bottom left panel of Fig.~\ref{fig:equilibrium_entropy}.
The overall trend is consistent with a power law, $\left\langle \Delta S\right\rangle \sim L^{-1.02}$.
This is comparable with the ideal scaling for intrinsic quantities
in Monte Carlo simulations, $\left\langle \Delta S\right\rangle \sim\frac{1}{\sqrt{L^{2}}}=L^{-1}$.

In the right panels, we show examples of real samples at several temperature
regimes, alongside samples synthesized by the trained generative model
at each temperature. The visual resemblance is striking, and it should
be immediately clear that this generative capability promises to be
a basis for powerful diagnostic techniques. In principle, the trained
model could now be used to perform a simulation. However, since knowledge
the distribution is sufficient information for evaluating any observable,
it is difficult to imagine scenarios where enough samples are available
to properly characterize the distribution, yet even more are still
wanted for some other purpose. One intriguing and promising approach
involves generation of systems that are too large to simulate \citep{kilgour_generating_2020,madanchi_simulations_2024,rotskoff_sampling_2024},
though it remains unknown whether this can capture, e.g., long-ranged
correlations.

\paragraph{Kinetic Ising model and dynamical phases.}

\begin{figure}
\includegraphics{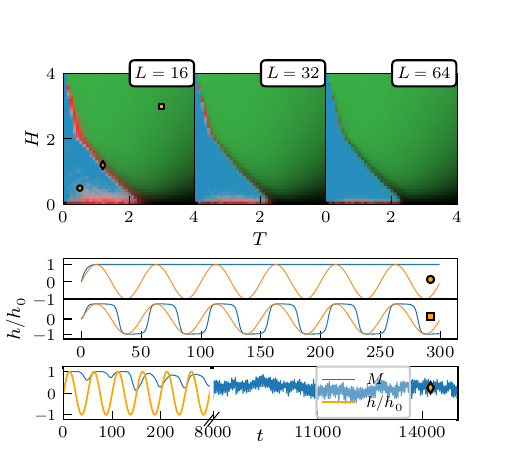}

\caption{(top panels) Phases of the kinetic Ising model at different oscillating
magnetic field amplitudes $H$ and temperatures $T$ and different
system sizes $L$. The order parameters characterizing the dynamically
ordered/ferromagnetic phase and the dynamically disordered/paramagnetic
phase are marked in blue and green, respectively. The parameter for
the strongly fluctuating/chaotic regime is marked in red (see text
for details). (middle and bottom panels) Traces of the magnetization
$\overline{M}$ at $L=16$ and in the three different regimes and
averaged over 96 realizations, with symbols marking the parameters
in the upper left panel.\label{fig:kinetic_ising_model}}
\end{figure}

Having established that modern density estimation techniques produce
accurate estimates of entropy in equilibrium, we note that nothing
in the method depends on taking samples from an equilibrium distribution.
We therefore now consider a nonequilibrium application. In particular,
we will explore the response of a 2D Ising ferromagnet to an oscillating
magnetic field $H\left(t\right)=H\sin\left(\omega_{0}t\right)$, as
described by Glauber dynamics \citep{glauber_timedependent_1963},
where every $L^{2}$ Metropolis steps are taken to be a time step.
This is a minimal toy model for exploring hysteresis in magnetic systems
\citep{zimmer_ising_1993,leung_response_1998,chakrabarti_dynamic_1999},
and is known to exhibit a dynamical phase transition \citep{lo_ising_1990,acharyya_response_1995,acharyya_nonequilibrium_1997,sides_kinetic_1998,korniss_dynamic_2000,yuksel_dynamic_2021}.
A dynamically disordered phase with zero period-averaged magnetization
emerges at weak magnetic fields, small field frequencies and high
temperatures; while a dynamically ordered phase with nonzero period-averaged
magnetic response is observed elsewhere \citep{rao_magnetic_1990,rao_hysteresis_1990}.

The dynamical phase diagram at several system sizes is shown in the
three top panels of Fig.~\ref{fig:kinetic_ising_model}, as a function
of the temperature $T$ and magnetic field amplitude $H$, and with
$\omega_{0}=\frac{2\pi}{50}$ throughout. To obtain this diagram,
we plotted the order parameter $O_{\text{Ordered}}=\left|\left\langle \overline{M}\right\rangle \right|$,
which characterizes the ordered phase, in blue (lower left regions);
and the parameter $O_{\text{Disordered}}=\left\langle \left|\overline{M}\right|\right\rangle \left(1-O_{\text{Ordered}}\right)$,
which characterizes the amplitude of oscillations in the disordered
phase, in green (upper right regions). Here, $\left|\cdots\right|$
is the absolute value, $\overline{\cdots}$ is an ensemble average
and $\left\langle \cdots\right\rangle $ is a time average over many
cycles. Examples of the time dependence of the ensemble-averaged magnetization
at $L=16$ in the ordered and disordered regimes, respectively, are
shown in the two middle panels. Similar phase diagrams have appeared
in the literature \citep{korniss_dynamic_2000,wang_phase_2012}, and
the kink in the phase boundary at $H=2$ was found to result from
the fact that the ordered regime actually comprises two dynamical
phases with different nucleation dynamics \citep{korniss_dynamic_2000}.

\paragraph{Strong fluctuations at the phase boundary.}

At certain parameters, both in and out of equilibrium, the Glauber
dynamics of small systems are highly stochastic. In particular, they
remain non-periodic for very long timescales even after averaging
over an ensemble of 96 realizations (see bottom panel of Fig.~\ref{fig:kinetic_ising_model}
for an example). This behavior has been observed and discussed in
the literature before \citep{lo_ising_1990,korniss_dynamic_2000}.
It can be characterized by the ensemble variance of the magnetization,
$O_{\text{Fluctuating}}=\left\langle \overline{M^{2}}-\overline{M}^{2}\right\rangle $,
which we plot in red in the top panels of Fig.~\ref{fig:kinetic_ising_model}.

Fluctuations of this kind are a mesoscopic effect: they become weaker
and appear over smaller parameter regimes as the system size grows
(i.e. going from leftmost to rightmost panel). Notably, $O_{\text{Fluctuating}}$
is large near equilibrium, at small magnetic fields and below the
critical temperature (horizontal red strips at bottom of three top
panels of Fig.~\ref{fig:kinetic_ising_model}). The hysteretic dynamics
there can be approximately described as stochastic hopping between
the two fully magnetized states, and is well understood \citep{sides_stochastic_1998,sides_kinetic_1999}.
We will focus on the other manifold of parameters where large fluctuations
appear: along the phase boundary (diagonal red strips in three top
panels).

\paragraph{Signature of fluctuations in the entropic limit cycle.}

The ensemble variance of magnetic fluctuations is easily accessible
from simulations. However, it is generally more difficult to access
experimentally than the mean magnetization: to obtain this, it would
be necessary to perform, e.g., many local measurements in a large
system, or many measurements of separate nanoscale crystals. Let us
therefore consider an alternative: characterizing the critical strongly
fluctuating regime using its entropy, e.g. by measuring the specific
heat globally for an ensemble of many nanoscale samples (or uncorrelated
regions in a large sample) at a series of temperatures. We will argue
that this is sufficient, with one complication.

Since the system occupies a very small subset of its possible states
in the near-equilibrium stochastic regime, we expect the entropy there
to be almost as low as that of the equilibrium ferromagnet. In contrast,
in the critical stochastic regime we expect the system to occupy a
wide variety of states at any given moment, and therefore exhibit
a high entropy compared to the ordered phase. On the other hand, the
disordered phase is also high in entropy, so it is not immediately
clear that these two scenarios can be distinguished from each other
by entropy alone.

\begin{figure}
\includegraphics{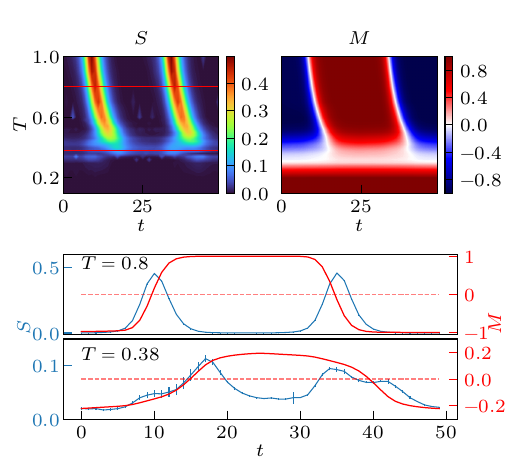}

\caption{(top left) Entropic limit cycle for system size $L=32$, at magnetic
field amplitude $H=2.5$. (top right) Magnetization limit cycle at
same parameters. (Bottom panels) Constant temperature cuts across
upper panels in the disordered (upper subpanel) and the critical strongly
fluctuating (lower subpanel) regimes, at the red horizontal lines
shown in top left panel.\label{fig:limit-cycle}}
\end{figure}

We therefore propose to study the time dependence $\overline{S}\left(t\right)$
of the entropy within a limit cycle rather than its time-averaged
value $\left\langle \overline{S}\right\rangle =\lim_{s\rightarrow\infty}\frac{\omega_{0}}{2\pi}\int_{s}^{s+\frac{2\pi}{\omega_{0}}}\mathrm{d}t\thinspace\overline{S}\left(t\right)$.
Experimentally, this could be measured by performing a global specific
heat measurement, but with heat injected by a sequence of short pulses
with the same periodicity as the field. The quantity $\overline{S}\left(t\right)$,
obtained from density estimation, is shown for the $L=32$ case in
the top left panel of Fig.~\ref{fig:limit-cycle} at a range of temperatures
spanning the transition. The corresponding limit cycle for the mean
magnetization is shown in the top right panel. The two panels below
show cuts across this data at two temperatures marked by horizontal
red lines in the top left panel.

In the disordered phase, as might be expected, the entropy is large
where the absolute value of the magnetization is small. As temperature
is lowered the magnetic and entropic response gradually slow down
and their peak shifts forward in time. At the other limit, where temperature
is near zero, the system enters the ordered phase and entropy is negligible.
Note that due to the constant initial phase of the field, at low temperature
we break symmetry in favor of positive magnetization (see also the
dynamics in the relevant panel of Fig.~\ref{fig:kinetic_ising_model}).
This manifests itself in the red band at the bottom of the top right
panel, and in the different magnitude of the two entropic peaks in
the lower panel.

At temperatures near the phase transition (see cut in lower panel
of Fig.~\ref{fig:limit-cycle}), magnetic response is weaker, while
entropic response is higher at all times. This directly demonstrates
that the system achieves low magnetization by exploring states that
are not fully magnetized, rather than by modifying the relative probability
of occupying the two opposite fully magnetized states. The converse
is true at lower temperatures (compare light red horizontal band in
top right panel to the corresponding region in the top left panel).
In that sense, the entropic limit cycle is a global measurement that
provides the kind of information that would only otherwise be available
in a local measurement (i.e. of the variance in the magnetization).

\paragraph{Conclusions.}

We showed that generative density estimation algorithms are capable
of accurately capturing the distributions characterizing certain statistical
mechanical systems, with no input other than samples obtained by traditional
simulation algorithms. After benchmarking the methodology on the equilibrium
Ising model, we used it to investigate nonequilibrium physics in a
kinetic Ising model. We obtained the time-dependent entropy of the
system when driven by a periodic oscillating magnetic field. We found
that global entropy can be used to demonstrate that near the dynamical
phase transition, in a regime characterized by strong local fluctuations,
the system attains a limit cycle where it is constantly exploring
a large manifold of states.

Within machine learning frameworks, generative density estimation
is straightforward to implement with a variety of algorithms. The
method shown here is widely applicable, since it requires only that
the system's configuration can be (at least approximately) mapped
onto a low-dimensional image. We've therefore provided a simple and
freely available code repository demonstrating how to reproduce and
extend our work \citep{gelman_pixelcnn_2024}.
\begin{acknowledgments}
G.C. acknowledges support by the Israel Science Foundation (Grants
No. 2902/21 and 218/19) and by the PAZY foundation (Grant No. 318/78).
\end{acknowledgments}

\bibliographystyle{apsrev4-2}
\addcontentsline{toc}{section}{\refname}\bibliography{library}

\end{document}